\documentclass[12pt]{article}
\usepackage{bm}
\usepackage{color}
\pdfoutput=1
\usepackage{jheppub}
\usepackage{amsfonts,amsbsy,amsmath,amssymb}
\usepackage{color}

\usepackage{amsfonts,amsbsy,latexsym,amssymb,amscd,amstext}
\usepackage{epsfig}
\usepackage{graphicx}

\usepackage{booktabs}
\usepackage{caption}
\usepackage{subcaption}
\usepackage{multirow}
\usepackage{algorithm}
\usepackage[noend]{algpseudocode}
\usepackage[utf8]{inputenc}

\newcommand{\be}{\begin{equation}}
\newcommand{\ee}{\end{equation}}
\newcommand{\ba}{\begin{array}}
\newcommand{\ea}{\end{array}}
\newcommand{\baa}{\begin{array}}
\newcommand{\eaa}{\end{array}}
\newcommand{\bea}{\begin{eqnarray}}
\newcommand{\eea}{\end{eqnarray}}

\newcommand{\Tr}{\mathrm{Tr}}

\newcommand{\ZMIN}{Z_{\mathrm{min}}}

\newcommand{\LMS}{\Lambda_{\overline{MS}}}

\title{The two-dimensional twisted reduced principal chiral model
revisited}

\author[a,b,c]{Antonio Gonz\'alez-Arroyo }
\author[d,e]{ and Masanori Okawa}

\affiliation[a]{ Instituto de F\'{\i}sica Te\'orica UAM/CSIC,  Nicol\'as
  Cabrera 13-15, \\
  Universidad Aut\'onoma de Madrid, E-28049 Madrid, Spain}
\affiliation[b]{
 Departamento de F\'{\i}sica Te\'orica,  M\'odulo 15,
 \\
  Universidad Aut\'onoma de Madrid, Cantoblanco, E-28049 Madrid, Spain}
 \affiliation[c] { Department of Physics, Faculty of Science, \\
Chulalongkorn University, Bangkok, Thailand
}
\affiliation[d]{  Graduate School of Science, Hiroshima University, \\
       Higashi-Hiroshima, Hiroshima 739-8526, Japan }
\affiliation[e]  {Core of Research for the Energetic Universe, Hiroshima University, \\
       Higashi-Hiroshima, Hiroshima 739-8526, Japan
}
\emailAdd{antonio.gonzalez-arroyo@uam.es}
\emailAdd{okawa@hiroshima-u.ac.jp}

\abstract{
Motivated by our previous study of the Twisted   Eguchi-Kawai model for non
minimal twists, we re-examined the behaviour of the reduced version of
the two dimensional principal chiral model. We show that this single
matrix model reproduces the same features as the standard lattice
model. In particular, scaling towards the continuum limit, the correct
value of the internal energy, the magnetic susceptibility and  the mass gap. Given our capacity to reach larger values of $N$, we use the reduced model to study  the nature and properties of its large $N$ phase transition existing at intermediate coupling. We conclude that the transition is of first order. }

\begin{document}
%\keywords{large N }
\preprint{%
{\flushright \vskip -0.5cm
\vbox{IFT-UAM/CSIC-18-55\\
FTUAM-2018-14 \\
HUPD-1803}
}}%

%\date{\today}

%\pacs{11.10.Lm,98.80.Cq}
%\keywords{}


\maketitle


%{\vskip 1cm}

\section{Introduction}
\label{s.intro}
The two-dimensional principal chiral model has several properties
in common with the more complex four dimensional gauge theories.
Thus, it provides an excellent laboratory to analyse  various  concepts and 
techniques applicable to both types of theories. Furthermore, the
model is integrable at the classical level, leading to an analytical
control of several properties of the quantum system. In particular,
this includes factorizable
S-matrix~\cite{Abdalla:1984iq}-\cite{Wiegmann:1984ec}, the determination
of the spectrum in the antisymmetric rank $r$ representation of the
SU(N) group, and the determination of the mass
gap~\cite{Balog:1992cm}. The latter is a nice example of the dimensional transmutation mechanism.
These theories have also been studied
numerically~\cite{Hasenbusch:1991aj}-\cite{Rossi:1993zc}-\cite{Rossi:1994yp}
on the lattice,  showing agreement with the analytical results and exhibiting 
precocious scaling when using the right bare parameters. 

Our motivation to address this study is related to the large $N$ limit of
the theory. As in the 4-d gauge theory case, this limit introduces certain simplifications without sacrificing its fascinating properties, which might be of help in attaining a full understanding of their dynamics. For example, the perturbative approach 
is restricted  to planar diagrams. Furthermore, the large $N$ limit at the
non-perturbative level can also give rise to new phenomena such as the
existence of large $N$ phase transitions. Indeed, the nature and
properties of the phase transition observed~\cite{Green:1981tr}-\cite{Campostrini:1994ih} in the lattice version of
the model at intermediate coupling has remained a subject of debate.

There are other open questions which have arisen in the more recent literature involving the principal chiral model and its large $N$ limit. This includes the study of form factors and possible exact results on correlation functions~\cite{Orland:2011rd}. Its role as a good testing ground also shows up in questions such as the existence and properties of other types of phase transitions~\cite{Narayanan:2008he}, which parallel  
 the observed critical behaviour of
Wilson loops in gauge theories in several space-time
dimensions~\cite{Narayanan:2006rf}-\cite{Narayanan:2007dv}.
In addition, for the same reason it also serves as a simplified situation in which to study phenomena such as resurgence~\cite{Cherman:2013yfa}. 

Our approach in this work is based in the ideas presented many years ago by Eguchi and Kawai~\cite{Eguchi:1982nm}. In particular, they proposed  that lattice Yang-Mills theories could become volume independent  in the large $N$ limit. This  led them to a matrix model, obtained by collapsing the whole lattice to a point, which 
was conjectured to be equivalent  to the ordinary theory in an infinite lattice. 
The phenomenon was called {\em reduction} and might be considered a
particular version of the so-called {\em volume independence}.
Although, the conjecture was soon shown to be false in the weak
coupling region of the model, several modifications were proposed to 
validate the reduction idea~\cite{Bhanot:1982sh}. The present authors put forward 
a modification of the original model, which goes under the name
Twisted Eguchi Kawai model
(TEK)~\cite{GonzalezArroyo:1982ub}-\cite{GonzalezArroyo:1982hz}. The idea is to introduce twisted
boundary conditions in the 1-point box. These conditions allow the
perturbative vacuum to respect a subgroup of the invariance group of
the original model, which is enough to guarantee reduction in the large
$N$ limit.

The reduction idea can be extended to non-gauge
systems~\cite{Parisi:1982gp}, and the present authors and others proposed a simple prescription 
to implement a similar twisted  reduction to other models~\cite{Eguchi:1982ta}-\cite{GonzalezArroyo:1982hz}. In particular this program was carried out soon afterwards for  the SU(N) principal chiral model~\cite{GonzalezArroyo:1984vx}-\cite{Das:1983jv}. 
We will refer to this model as the twisted reduced principal chiral model (TRPCM). Apart from analysing the
Schwinger-Dyson equations of the ordinary and reduced model and identifying the necessary conditions for reduction to apply, the authors studied the model numerically in both two and four dimensions.
From the beginning it was clear that the two-dimensional model was
rather tricky. In particular, the ordinary model has a continuous
global symmetry which cannot be broken in two-dimensions, while the 
reduced model is only invariant under a discrete symmetry. Results 
were however not incompatible with reduction working in both
dimensions. 

Many years later the twisted prescription was realized to be a discrete version of non-commutative field theories~\cite{Ambjorn:1999ts}. Indeed, the lagrangian and Feynman rules for these theories appeared first when constructing a continuum version of the twisted reduced models~\cite{GonzalezArroyo:1983ac}. The connection is also there for the principal chiral model~\cite{Profumo:2001hm}. 
With this renewed interest as an additional motivation, the TRPCM was then re-analysed  in Ref.~\cite{Profumo:2002cm}
using the more powerful computer resources and methodologies available at the time. The validity of the equivalence between the reduced model and the infinite volume lattice model appeared to be in question for the particular case of the two-dimensional model at sufficiently small values of the 
lattice 't Hooft coupling. A
particularly relevant  quantity was the susceptibility which was seen to grow with $N$ for the reduced model instead of converging to the value of the 
ordinary model. 

Indeed, this was the first of a series of problems reported by various authors questioning the validity of the twisted reduction idea. In particular, the four-dimensional gauge model, the TEK model,  showed signals of $Z_N$ center symmetry-breaking at large enough values of $N$ and intermediate values of the lattice coupling~\cite{Ishikawa}-\cite{Bietenholz:2006cz}-\cite{Teper:2006sp}-\cite{Azeyanagi:2007su}. The breakdown of the symmetry invalidates the non-perturbative proof of reduction. 

In view of these conflicting results, the present
authors~\cite{GonzalezArroyo:2010ss} observed that all of the problems arose when the ratio of the discrete flux $k$ over $N$ became smaller than a certain value ($\sim0.1$). This can arise when the entropy of certain symmetry-breaking vacua can overcome their higher energy and dominate the path integral. In other cases it is rather the ratio of the modular inverse of the flux $\bar{k}$ over $N$, controlling the suppression of non-planar diagrams,  which becomes too  
small. The  proposal made in Ref.~\cite{GonzalezArroyo:2010ss} is to take the large $N$ limit keeping the ratios $k/N$ and $\bar{k}/N$ bigger than a certain threshold estimated to be around $0.1$. With this additional condition a detailed verification of the equivalence has been carried out for the gauge theory\cite{Gonzalez-Arroyo:2014dua}. Not only there is no sign of symmetry breaking but also in some quantities the agreement between the observables has been tested up to the fifth decimal place. 
The successful comparison extends also to quantities in the continuum limit, such as the string tension~\cite{agaokawa3}.  

A deeper understanding has followed from our study of the 2+1 dimensional gauge theory~\cite{Perez:2013dra}-\cite{Perez:2014sqa}. In that case the flux choice affects the existence or not of tachyonic instabilities. Our results indicate that the main quantity to control is $\ZMIN$ defined as follows
\be
\label{zmin}
\ZMIN=\min_{e}  e ||\frac{ke}{N}||
\ee
where $e$ is an integer coprime with $N$, and the symbol $||\cdot||$ denotes distance to the nearest integer. To avoid instabilities one must choose  $\ZMIN$ to be larger that $0.1$~\cite{Chamizo:2016msz}. Notice that by definition $\ZMIN$ is smaller than $k/N$ and $\bar{k}/N$, so that the new condition implies the previous ones. A more detailed analysis of these questions and its implication can be seen in Ref.~\cite{newtwoplusone}.

It is the purpose of this paper to apply these ideas to the TRPCM, to see if they are able to circumvent the problems 
mentioned earlier. The main restriction is to make an adequate choice  of the flux $k$ characterizing twisted boundary conditions on the 2-torus.  The choice does not alter neither the general proof of reduction based on Schwinger-Dyson equations, nor the equivalence in perturbation theory. Furthermore, it comes without any   additional  computational cost. As we will see our results follow the same pattern that was observed for the gauge theories, restoring the validity of the reduction idea  in this domain.

The outline of the paper is the following. In section 2, we review some of the main properties of the principal chiral model, its lattice version  and the
twisted reduced version. In section 3 we revisit the study made by
Profumo and Vicari~\cite{Profumo:2002cm}, generalizing their results obtained for $k=1$ to arbitrary flux.
This illustrates how the problems reported only occur in the same {\em dangerous} region as for the reduced gauge theory.
In section 4 we made a direct test of the validity of reduction by comparing the value 
of various observables in the large volume and large $N$ ordinary lattice model with those obtained for the TRPCM with an adequate choice of flux. 
The agreement in some quantities is less than a permille. In section 5 we study the mass gap for the reduced model and show that it satisfies the predicted  scaling behaviour towards its
continuum limit. This shows that the reduction extends to the continuum limit of both theories, an important ingredient to show its usefulness. Next, in section 6 we use  the ability  of the reduced model to explore larger values of $N$ to investigate the nature of the large $N$ phase transition mentioned earlier. This is an instance in which things only become clear for rather large values of $N$.  Finally, in the last section we present our
conclusions.  

\section{The Reduced Principal Chiral Model}
\subsection{Short review of the two dimensional principal chiral model}
The SU(N) principal chiral model is a quantum field theory whose lagrangian
density is given by
\be
{\cal L}= \frac{1}{g^2} \mathrm{Tr}(\partial_\mu U(x) \partial_\mu
U^\dagger(x))
\ee
where the field $U(x)$ takes values in the fundamental representation
of the SU(N) group. 
In two-dimensions the theory is asymptotically free and generates a
mass gap by dimensional transmutation~\cite{Polyakov:1977vm}. The theory is invariant under 
an (SU(N)$\times$SU(N))/Z$_N$ global symmetry
\be
U(x) \longrightarrow \Omega' U(x) \Omega^\dagger
\ee
which in two-dimensions cannot be broken by the Mermin-Wagner-Coleman
theorem~\cite{Mermin:1966fe}-\cite{Coleman:1973ci}. 

By using different techniques several properties of the model have
been established. In particular it follows from the S-matrix structure
that the system contains bound states having the quantum numbers of
the r-antisymmetric representation of SU(N), whose masses $m_r$ follow the 
pattern
\be
m_r = m \frac{\sin(\pi r/N)}{\sin(\pi/N)} .
\ee
The mass gap itself $m$ can be deduced by applying a magnetic field
coupled to the suitable chosen charge, and combining  S-matrix
results with perturbation theory:
\be
\frac{m}{\LMS} =\sqrt{\frac{8\pi}{e}}\, \frac{N \sin(\pi/N)}{\pi}
\ee
where $\LMS$ is the Lambda-parameter of the theory,
defined as for Yang-Mills theory. 

\subsection{The lattice version of the PCM}
Formulation of the model on the lattice allows to study these
properties and others in a non-perturbative fashion (For a review of the main results and a list of references we address the reader to Ref.~\cite{Rossi:1996hs}). The partition function   is given by
\be
Z= \int \prod_{n} dU(n) \exp\{-b N\sum_{n} \sum_\mu
 \mathrm{Tr}(\delta_\mu U(n) \delta_\mu U^\dagger(n))\}
\ee
where $\delta_\mu U(n)=U(n+\hat{\mu})-U(n)$ is the discretized
derivative. The coupling $b=1/(g^2_0 N)$ is the inverse of the lattice `t
Hooft coupling. 

The main observable in the lattice  model
is the correlation function $G(n)$
\begin{equation}
 G(n)= \frac{1}{N} \langle \Tr(U(0) U^\dagger(n)) \rangle
\end{equation}
where $n$ is a two-component integer vector. In particular one sees that the internal energy of the model can be written as follows
\be 
E=1-\frac{1}{2}\langle \mathrm{Re}(G(1,0)+G(0,1)) \rangle .
\ee

It is convenient to consider the correlation function projected onto   zero spatial momentum given by
\begin{equation}
\bar{G}(t)=\sum_n G(t,n)
\end{equation}
where $t$ is an integer. This function falls off at large $t$ exponentially with $t$. Its coefficient defines the lattice mass gap $M$, which is a function of $b$ and $N$.
Another interesting observable is the susceptibility $\chi$ which is the sum over $t$ of $\bar{G}(t)$.

In order to study the continuum limit we can use the first two coefficients of the
beta-function~\cite{McKane:1979cm} 
($\beta_0=\frac{N}{8 \pi}$, $\beta_1=\frac{N^2}{128 \pi^2}$) to define the lattice spacing by the formula:
\be
\label{scaling}
a \Lambda_L =\sqrt{8 \pi b}\, e^{-8 \pi b} .
\ee
Computing quantities in perturbation theory for large $b$ and comparing with the results obtained using dimensional regularization one obtains the relation~\cite{Shigemitsu:1981gi}:
\be
\frac{\LMS}{\Lambda_L}=\sqrt{32} \exp\{\frac{\pi(N^2-2)}{2 N^2} \} .
\ee

One can also take a particular short distance observable and use it to define a different bare coupling constant. For example by using the internal energy one can define
\be
b_E=\frac{N^2-1}{8 N^2 E}
\ee
which makes the leading order perturbative formula for the internal energy exact, when written in terms of this coupling. This choice first proposed by Parisi has been used extensively in Yang-Mills theory, and also in the principal chiral model~\cite{Rossi:1993zc}. One can improve the scale determination by computing the next-to-next to leading coefficient of the beta function in this so-called $b_E$ scheme. This gives the formula~\cite{Rossi:1994yp}
\be 
\label{scalingP}
a_E \Lambda_E =F(b_E)\equiv \sqrt{8 \pi b_E}\, e^{-8 \pi b_E} (1-\frac{0.00884}{b_E}) .
\ee 
Using perturbation theory one can obtain the relation between $\Lambda_E$ and $\Lambda_L$ as follows:
\be 
\frac{\Lambda_E}{\Lambda_L}= \exp\{\frac{\pi (N^2-2)}{4N^2} \} .
\ee
Monte Carlo simulations (and strong coupling expansions) for various $N$ and various $b$ showed that the lattice model Green function approaches the behaviour of the continuum theory and that scaling violations are small when computing in the $b_E$ scheme~\cite{Rossi:1994yp}. 

Here we will focus on the large $N$ version of the lattice model which has been
subject of interest since very early times~\cite{Green:1981tr}. In particular, analysis of
the strong coupling expansion~\cite{Campostrini:1994kf} and Monte
Carlo simulations~\cite{Campostrini:1994ih} indicate the
presence of large $N$ phase transition around $b=0.306$ displaying a peak in
the specific heat. The nature of the phase transition and the corresponding exponents are not completely settled. 
We will address this problem in sect. \ref{sect:6}.

\subsection{The reduced principal chiral model}
The twisted reduced principal chiral model~\cite{GonzalezArroyo:1984vx}-\cite{Das:1983jv} is a one  matrix model whose
large $N$ limit is proposed to  be equivalent to the standard lattice version of
the principal chiral model. The action is obtained by the replacement 
of the spatial displacement by the adjoint action by  certain SU(N)
matrices:
\be
\delta_\mu U \longrightarrow  \Delta_\mu U \equiv \Gamma_\mu U \Gamma_\mu^\dagger -U
\ee
and then dropping the space dependence. 
In this way the partition function becomes
\be
Z=\int dU\ \exp\{-bN \sum_\mu \mathrm{Tr}(\Delta_\mu U \Delta_\mu
U^\dagger)\} .
\ee
Notice that the only dynamical degree of freedom is the SU(N) matrix
$U$. In two dimensions the $\Gamma_\mu$ matrices are forced to satisfy
the constraint
\be
\Gamma_1 \Gamma_2= \exp\{\frac{2 \pi i k}{N} \} \Gamma_2 \Gamma_1 .
\ee
If $k$ is chosen co-prime with $N$ the solution is unique modulo equivalences (global gauge transformations). The restriction to coprime values is also important in perturbation theory. A particular solution (choice of basis) can be given by 't Hooft shift and clock matrices $P_{ij}=\delta_{j\ i+1}$ and $Q={\rm diag}(1,z,z^2,\cdots ,z^{N-1})$ with $z=\exp\{\frac{2 \pi i}{N} \}$.
They satisfy $PQ=zQP$.  Then $\Gamma_1=P$ and $\Gamma_2=Q^k$.   Notice, however, that for even $N$ some of the matrices have determinant $-1$. Thus, if we impose that $\Gamma_\mu$ belong to SU(N) one should rather take $\Gamma_1 = z^{1/2} P$ and $\Gamma_2 = (z^{1/2} Q)^k$.

Consistently with the association of displacement with adjoint action one can also define the corresponding correlator to be 
\be
G_R(n)=\frac{1}{N} \langle \Tr(U\Gamma(n)U^\dagger \Gamma^\dagger(n)) \rangle
\ee
where with the previous choice of basis $\Gamma(n)$ is given by 
\be
\Gamma(n)=P^{n_1} Q^{k n_2} .
\ee
The internal energy is written in terms of $G_R$ as in the ordinary model.

In the aforementioned basis the zero-momentum projected correlation function depends only on the diagonal elements 
of $U$ as follows
\be 
\bar{G}_R(t)=\sum_l \langle U^*_{t+l\, t+l} U_{l l} \rangle .
\ee
Then summing over $t$ we obtain the susceptibility
\be 
\chi= \langle |\Tr(U)|^2 \rangle .
\ee

\section{Flux dependence of physical quantities}

\begin{figure}
\centering
\begin{subfigure}[b]{0.5\textwidth}
\includegraphics[width=\textwidth]{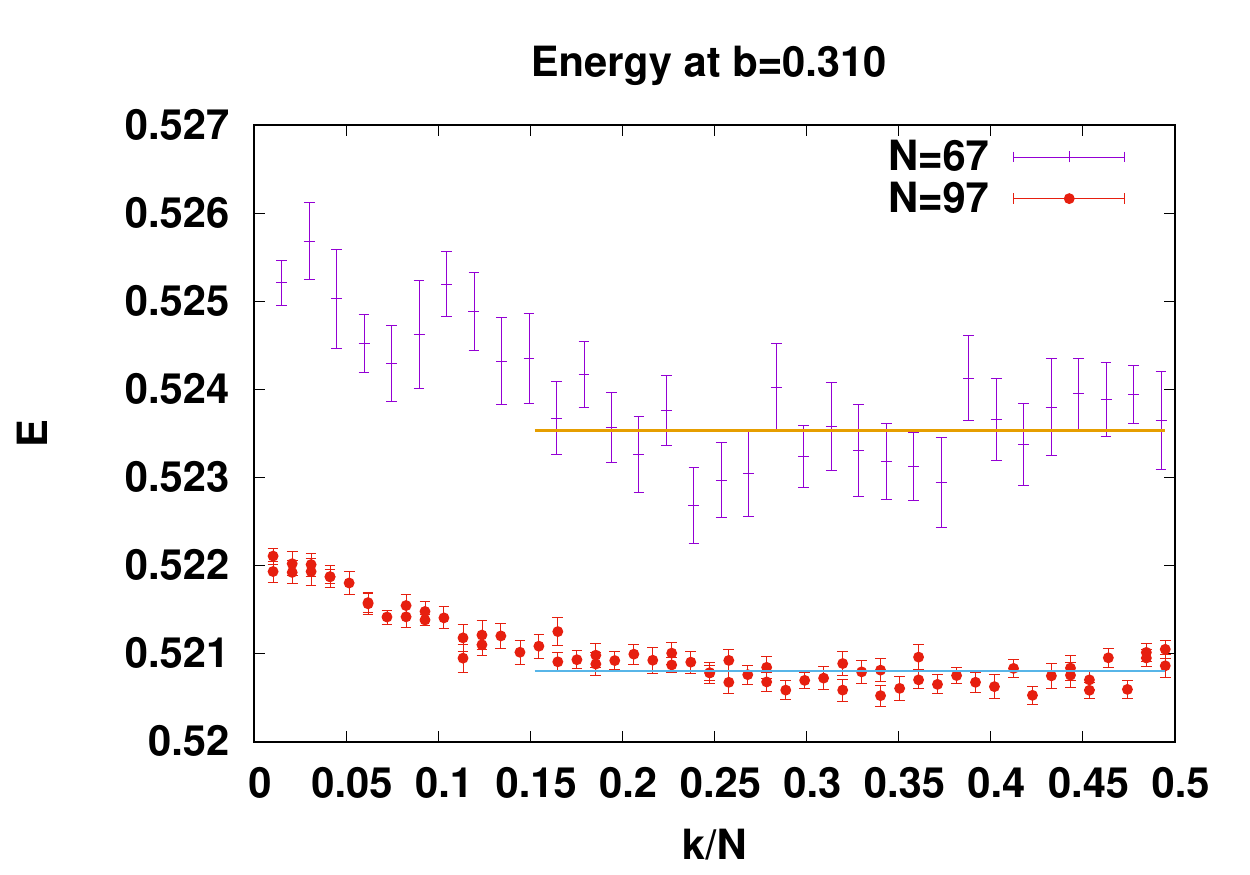}
\caption{ Internal Energy as a function of $k/N$.}
\label{fig:1a}
\end{subfigure}%
\begin{subfigure}[b]{0.5\textwidth}
\includegraphics[width=\textwidth]{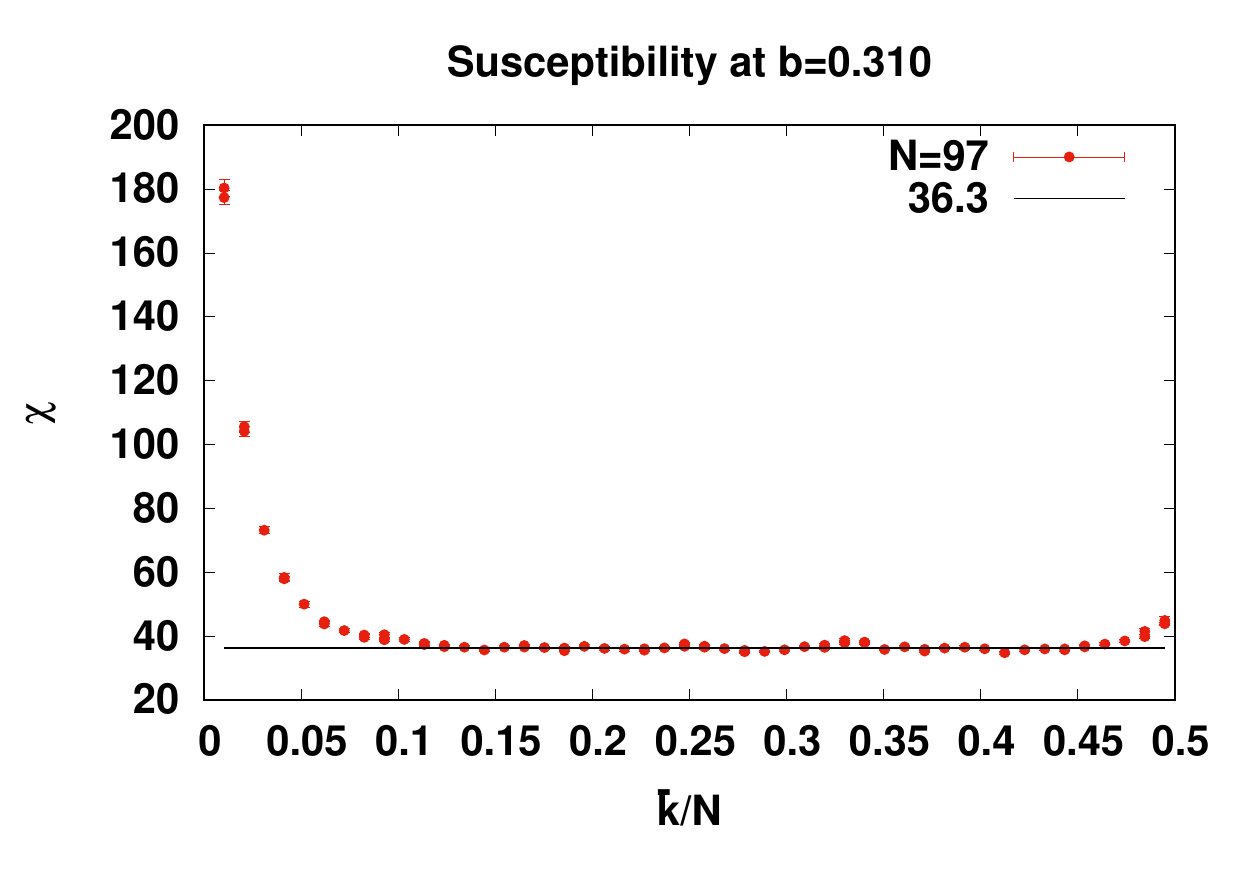}
\caption{Susceptibility as a function of $\bar{k}/N$.}
\label{fig:1b}
        \end{subfigure}
\caption{Flux dependence of physical quantities at $b=0.31$.}
\label{fig:1}
\end{figure}

The results obtained by several authors for the TEK model and for the 2+1 dimensional gauge  model have shown the importance of choosing the flux appropriately. In this section we will show that some physical  quantities depend strongly on the value of the integer $k$. Since our original motivation is to study the problems reported by  Profumo and Vicari~\cite{Profumo:2002cm} (corresponding to $k=1$), we have focused in the same quantities studied by them: the internal energy  $E$ and the susceptibility $\chi$. For the same reason we concentrated on the value $b=0.31$. At this coupling value we  analyzed several prime values of $N$ ranging from 23 to 137, and studied all possible non-zero values of the flux $k$ for them. 
Simulations of TRPCM have been done with both heat bath\cite{Fabricius} and over-relaxation methods\cite{overrelaxation}. In Fig~\ref{fig:1} we illustrate our results. On the left figure we show the results for the energy $E$ at b=0.31 and two values of $N$ (67 and 97). We see a smooth dependence of the result on $k/N$, peaking for small values and showing a plateau above $k/N \ge 0.15$. The pattern is general and also holds for the other values of $N$, not displayed. Two other features are that the value decreases with $N$ (we will study that later) and also that the spread of values decreases when $N$ grows. The results are compatible with the findings of Ref.~\cite{Profumo:2002cm} which correspond to the extreme left value of each set of data ($k=1$). In any case, despite the slight differences, the authors of Ref.~\cite{Profumo:2002cm} concluded that there was no problem with the  energy of the reduced model, since it was apparently converging to the same value  as the ordinary model $\sim 0.519$. 

However, the situation changed completely when they studied  the susceptibility $\chi$. The value given by the reduced model differed considerably from  the one of the large $N$ principal chiral model. Furthermore, the  difference kept increasing  with growing values of $N$. 

Our results for $N=97$  are displayed in the right subfigure of Fig.~\ref{fig:1}. The susceptibility is plotted as a function of $\bar{k}/N$. The data shows a strong peak for small values of this quantity. However, for values larger than $\sim 0.12$ most of the data are consistent with each other and a value $\chi=36.3$ drawn as a horizontal line in the same plot. Note that the value $k=\bar{k}=1$ used in Ref.~\cite{Profumo:2002cm} corresponds to the leftmost value of the x-axis where the peak is more pronounced.  

The pattern repeats itself for other values of $N$. As $N$ increases the left  peak becomes narrower but at the value  $k=\bar{k}=1$ actually increases. This is consistent with the observation of Ref.~\cite{Profumo:2002cm}. For completeness we give the values obtained for all our values of $N$. The susceptibility at $k=\bar{k}=1$ takes the values 50, 91, 102, 127, 137, 172, 180 and 226 for the sequence of values $N$=23, 43, 53, 61, 67, 89, 97 and 137. 
The growth seems to go as $N$ at smaller values, which  then reduces to $\sim N^{1/2}$.

What is the origin of the peak? For the purpose of answering the question, we analyzed the correlator at the maximum distance $N/2$: $\bar{G}_R(N/2)$. For a gapped theory it should go to zero when $N$ grows. A non-zero asymptotic value reflects  a non-vanishing expectation value. With the standard definition a non-zero expectation value gives a susceptibility diverging with the volume (in our case $N^2$).  It is possible to subtract this constant piece and define a new susceptibility. Indeed, if we take our result and subtract 
$ C N^2  \bar{G}_R(N/2)$ with $1<C<1.025$ we eliminate the peak for all values of $N$. 

Our prescription is, however, much simpler and compliant with the one adopted for the 2+1 and 3+1 dimensional gauge theories~\cite{GonzalezArroyo:2010ss}-\cite{Perez:2013dra}. As explained in the introduction one should take the continuum limit keeping both $k/N$ and $\bar{k}/N$ bigger than a certain threshold which is roughly equal to 0.1. Indeed, a better way to select the adequate values of $k$ is to constrain the quantity $\ZMIN(N,k)$, defined by Eq.~\eqref{zmin}, to be larger than that same value. 
A good test that this criteria eliminates the problems associated with the non-zero expectation value is to plot $\bar{G}_R(N/2)$ as a function of $N$. Our data give a neat exponential fall-off of the form $0.278 \exp{(-0.176 N/2)}$ until the moment that this value is smaller than the statistical errors ($N\sim 97$). The fall-off is the characteristic one for a gapped theory with no expectation value.  

In conclusion, we emphasize that, also for the two-dimensional principal chiral model,  it is convenient to adopt the same criteria for flux selection as advocated in Ref.~\cite{GonzalezArroyo:2010ss}. As mentioned in the introduction, for the  2+1 dimensional case~\cite{Chamizo:2016msz}-\cite{newtwoplusone} we observed that this choice avoids the presence of symmetry-breaking phase transitions at $N$ large. However, even if no transition is present it is still very important for practical purposes  to reduce the size of the finite $N$ corrections. In the following sections we will adopt this criteria and we will explore the behaviour of the reduced model at large $N$ both on the lattice and in the continuum limit.

\section{Validity of reduction at large $N$}

In this section we make a precision study of the validity of the volume independence hypothesis for the case of the two-dimensional principal chiral model. A successful test
was already done in    Ref.~\cite{Profumo:2002cm}  for 3 and 4 dimensions. 
The test for the 2-dimensional model failed for the reasons explained in
the previous section. Having identified the source of the problem we will 
retake the original goal. Notice, that it is not only a question of proving 
the validity but also a way to estimate the errors associated to finite $N$.
Our study is very similar to the one done by the present authors~\cite{Gonzalez-Arroyo:2014dua} for Wilson loops on the TEK model.  

\begin{figure}
\centering
\begin{subfigure}[b]{0.5\textwidth}
\includegraphics[width=\textwidth]{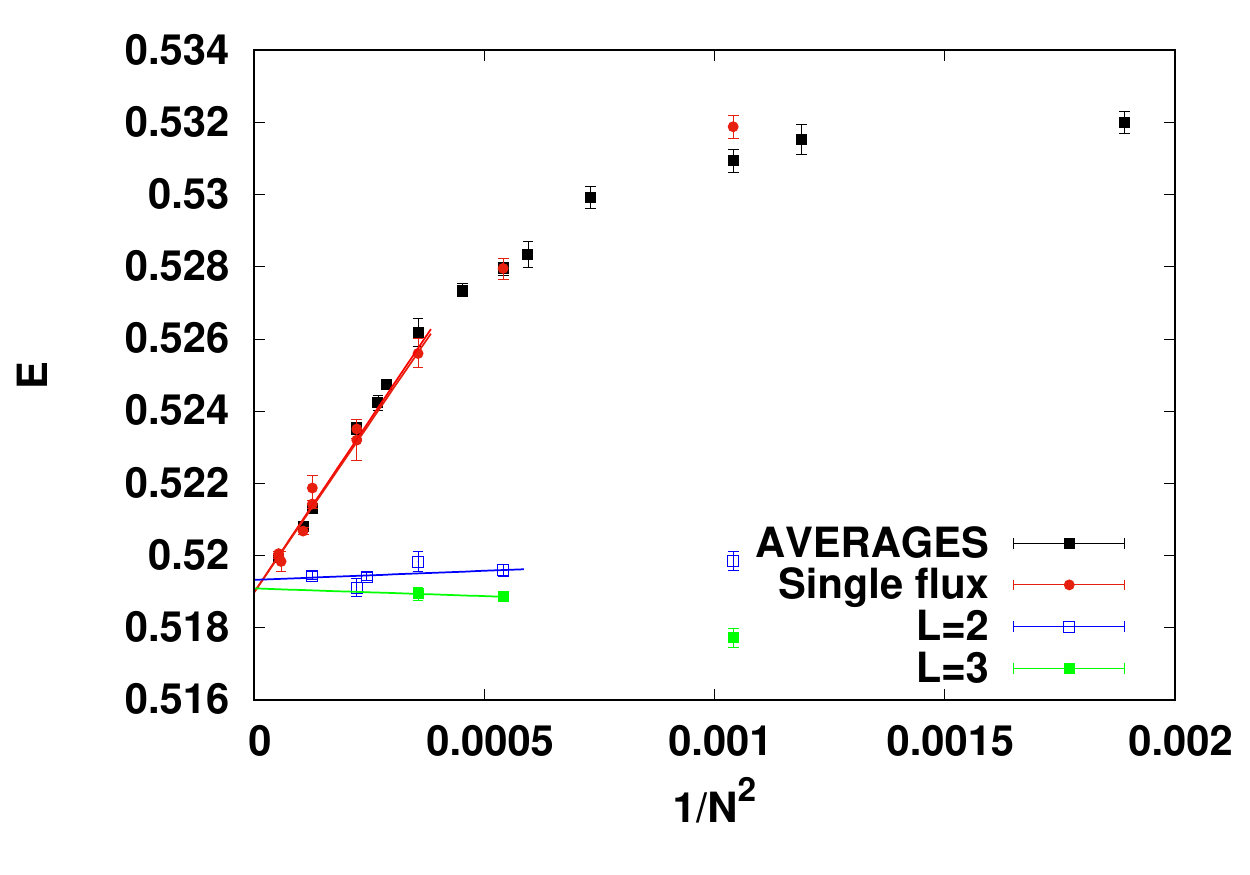}
\caption{ Internal Energy. }
\label{fig:2a}
\end{subfigure}%
\begin{subfigure}[b]{0.5\textwidth}
\includegraphics[width=\textwidth] {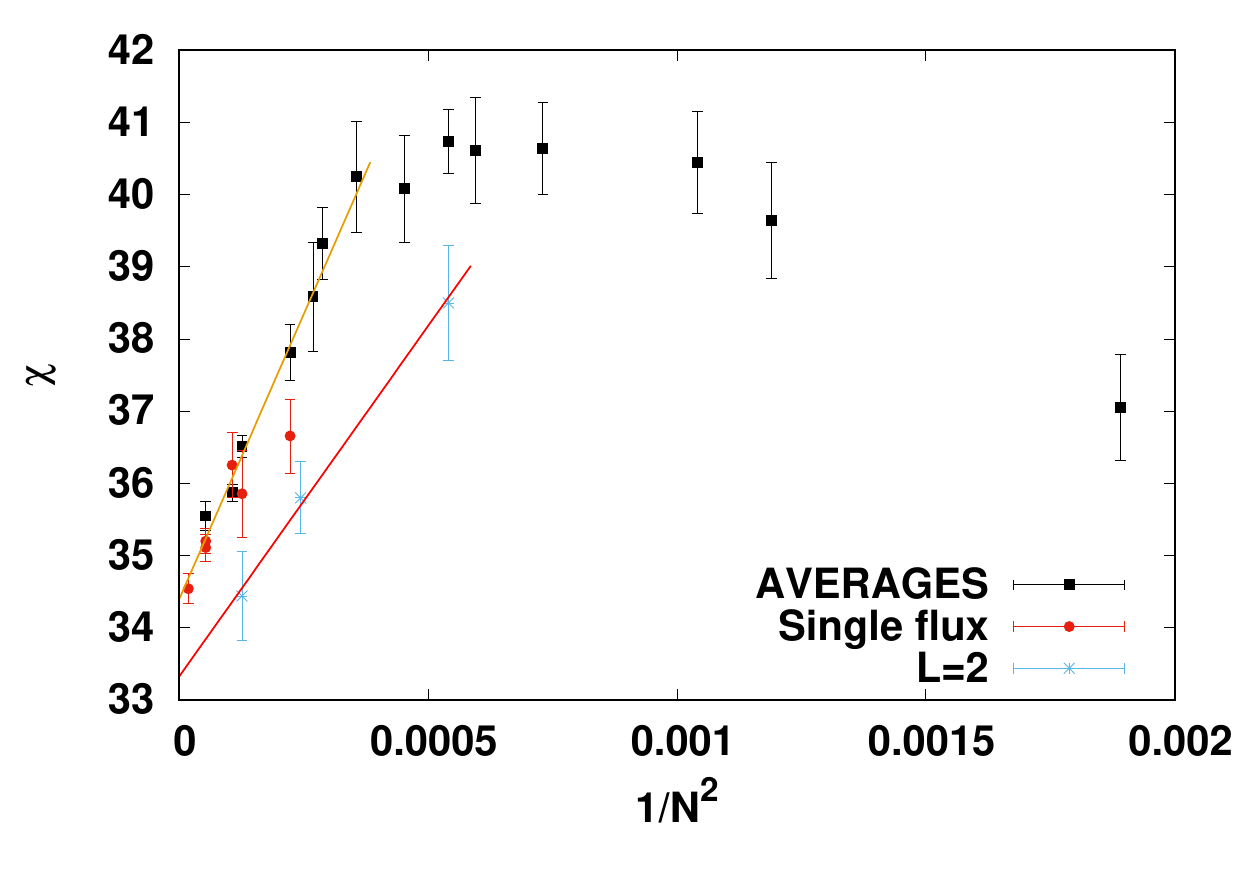}
\caption{Susceptibility. }
\label{fig:2b}
\end{subfigure}%
\caption{$N$ dependence of physical quantities at $b=0.31$. Points labeled {\em AVERAGES} are obtained for the TRPCM averaging over various values of $k$. The single flux correspond to other measurements at fixed value of $k$. The points labeled $L=2,3$ are obtained for an $L\times L$ lattice with twisted boundary conditions.}
\label{fig:2}
\end{figure}

Sticking first  to the value b=0.31 of the lattice coupling, we summarize our results in Fig.~\ref{fig:2}. The left plot focuses on the internal energy $E$ and the right one on the susceptibility $\chi$. The data points labelled as {\em AVERAGES} correspond to the results mentioned on the previous section averaged over all values of the flux $k$ satisfying the criteria $\ZMIN(N,k)\ge 0.15$. We have also added other measurements performed at single flux values but satisfying the same criteria. The $N$-dependence of the results shows two regimes. One up to values of the order of $N=50$ which is rather flat, and one for larger values which is consistent with a $1/N^2$ linear dependence. Extrapolating the results to infinite $N$ we get $E_\infty=0.5190(2)$ and $\chi_\infty=34.4(2)$. These values are consistent with the values obtained by  previous authors from extrapolation in the ordinary lattice model (0.519 and 34.1(2) respectively). We have also made our own  high-precision simulation in a $64\times 64$ lattice at $N=64$ and periodic boundary conditions. Our results are $E=0.519004(13)$ and $\chi=34.67(4)$.  
Thus, within the precision of our results (a permille and a percent respectively) we have verified that the reduced model tends to the infinite volume lattice model at large $N$. 

We have also studied the coupling value $b=0.32$. In that case we did not scan all values of $k$ but simply chose appropriate ones according to our criteria. In particular we studied several values of $N$ (43, 53, 67, 89, 97, 131, 137, 233). The results  of the energy could be fitted (with good chi square) by the function  $E=0.485758(15)+2.7(2)/N^2$.  This matches with the rough estimate $0.485$ given in Ref.~\cite{Profumo:2002cm}.  The situation for the susceptibility is not that good. A good fit to a   linear function in  $1/N^2$ is only possible for $N > 67$. The extrapolated result at $N=\infty$ is 63.4(6) which is not too far from  the estimate of 65 provided earlier\cite{Profumo:2002cm}. The coefficient of $1/N^2$ being huge ($\sim 120000$). Although our  results for the reduced model do not seem to be in conflict with those of the ordinary lattice model,  a more precise comparison would require   more statistics  for the ordinary model simulation.

As mentioned in Ref.~\cite{Profumo:2002cm} the $N$-dependence is rather strong compared with that of the ordinary model. To a large extent this is because the $N$-dependence takes the role of the volume dependence which for the ordinary model is much stronger. For example, the value of the energy for the ordinary model with $N=43$ on a $6\times 6$ lattice with periodic boundary conditions is   0.4889, which is much further away from the infinite volume result than any of twisted results. 
In any case, it is clear that taking twisted boundary conditions pays off. For example, one can study a model reduced to a $2 \times 2$ box with twisted boundary conditions (a matrix model of 4 matrices instead of 1). In that case the corresponding values of the energy and susceptibility also appear in the plot under the label $L=2$. The plot also contains  (under the label $L=3$)  the results obtained on a $3 \times 3 $ lattice with twisted boundary conditions. For the energy the $L=2$ data are independent on $N$ within errors. This is in part accidental, as there is a compensation of finite volume errors which tend to raise the value with the genuine finite $N$ corrections of the ordinary SU(N) model which tend to decrease it. In particular, we mention that the results at $N=64$ but using twisted boundary conditions with $k=19$ on a $2\times 2$ lattice gives $E=0.51946(5)$ and $\chi=35.8(5)$, and in a $3\times 3$ lattice
$E=0.51918(10)$ and $\chi=35.94(35)$. At $b=0.32$ we also 
studied the twisted model on a $2\times 2$ lattice and N=67. The energy becomes $E=0.48569(5)$ and the susceptibility becomes $\chi=70.2(9)$.

Our comment about the "partially reduced" model could be of practical importance for simulations. The larger number of degrees of freedom might pay off. This also neutralizes a similar criticism presented  by Profumo and Vicari. For example, they studied a lower value of $b$ ($b=0.28$). In that case, they observed convergence of the reduced model results towards the ordinary model at large $N$. However, they observed that in the coefficient of the $1/N^2$ corrections was two orders of magnitude larger for the reduced model. For example, they obtained at $N=30$ a value of the susceptibility of $9.49(3)$ to be compared 
with the large $N$ result estimated to be 7.02(2). We have performed the measurement on a $2\times 2$ twisted box with flux $k=11$ and we obtained $7.09(5)$. The correction then is then not much bigger than the one of the ordinary model.

\section{Continuum limit}
\begin{figure}
\centering
\begin{subfigure}[b]{0.5\textwidth}
\includegraphics[width=\textwidth]{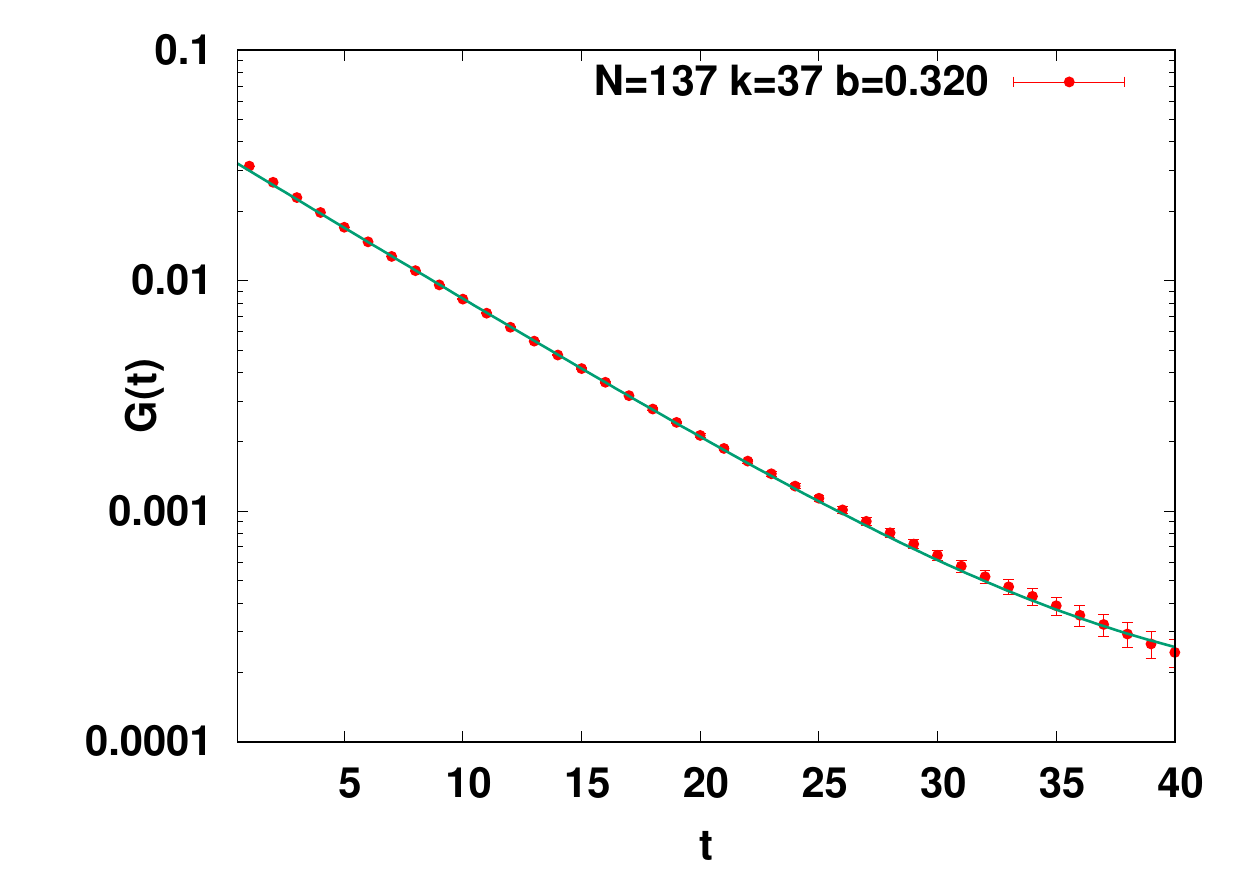}
\caption{ Correlator $\bar{G}_R(t)$. }
\label{fig:3a}
\end{subfigure}%
\begin{subfigure}[b]{0.5\textwidth}
\includegraphics[width=\textwidth]{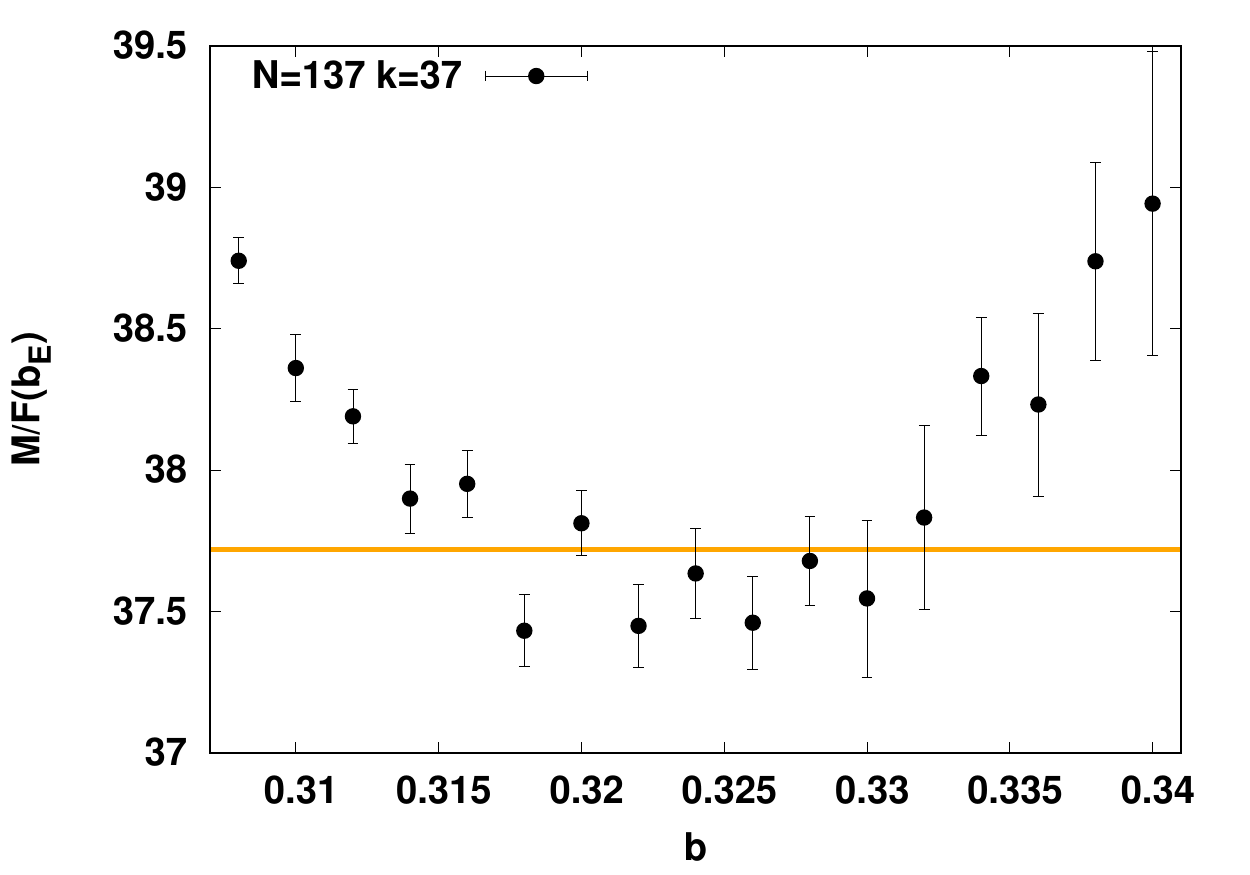}
\caption{Determination of $m/\Lambda_E$. The horizontal line is the continuum prediction.}
\label{fig:3b}
        \end{subfigure}
\caption{Analysis of the correlator and scaling.}
\label{fig:3}
\end{figure}

In this section we will study the reduced model in the continuum limit and
compare it with the exact results that we have for  this  model. To be precise we will concentrate on the calculation of the mass gap $m$ extracted from  the zero-momentum projection of the propagator $\bar{G}_R(t)$. For that purpose we simulated the reduced model for various values of $b$ ranging from $b=0.3$ up to $0.37$ and various values of $N$ (67,89, 97 and 137). We will concentrate here on the $N=137$ data for which we have a much higher statistics. In that case we generated 120000 configurations
separated by 100 sweeps between each two. We did this for two different fluxes $k=37$ and $50$. For each configuration we computed the zero-momentum correlator $\bar{G}_R(t)$ with errors estimated by jacknife. As a sample of the quality of the data we show in Fig.~\ref{fig:3a} the correlator for $b=0.32$ and $k=37$. The data is then fitted to a 3 parameter formula as follows:
\begin{equation}
A+B \exp(-M t) .
\end{equation}
In this way we obtain a lattice mass $M$ as a function of $b$. Using the fitting range $5\le t\le 25$ we obtain good fits with chi square per degree of freedom smaller than 1. The corresponding fit for $b=0.32$ is shown as a solid line in  Fig.~\ref{fig:3a}. 

According to the scaling analysis explained earlier, we consider that our lattice mass $M$ equals $m a_E(b)$, where $a_E$ is given in Eq.~\eqref{scalingP}. Substituting we obtain 
\be
M=\frac{m}{\Lambda_E} F(b_E)
\ee
which according to scaling should hold asymptotically for large $b$. In Fig.~\ref{fig:3b} we plot the ratio $M/F(b_E)$ obtained from our data in the weak coupling region ($b>0.306$; The $b_E$ scheme is singular below). The result is remarkably constant to within $3\%$ in the whole range. This is quite non-trivial as $F(b_E)$ is three times larger in lower edge than in the upper edge of the displayed region. Furthermore, the constant is predicted to be 
\be 
\frac{m}{\Lambda_E}=\frac{m}{\LMS}\ \frac{\LMS}{\Lambda_L}\ \frac{\Lambda_L}{\Lambda_E}= 16 \sqrt{\frac{\pi}{e}}\  \frac{\sin(\pi/N)}{\pi/N}\, \exp\{\pi\frac{N^2 -1}{4 N^2} \}= 37.72 .
\ee
This value is displayed as the orange line in the figure. The agreement is remarkable. Nonetheless, it must be mentioned that  our result has systematic errors which are estimated to be of order $5\%$. This comes from changes in the fitting range and the choice of $k$. Smaller values of $N$ give similar results but the plateau is shorter.

A few comments on our result are necessary. The first concerns the constant term $A$ in the fitting formula. This term is needed to get high precision fits for $b>0.31$. This constant term actually grows as $b$ grows. Comparing similar fits for smaller values of $N$, one sees  that the constant decreases with $N$. This is presumably also related with the non-zero value of $\Tr(U)$ observed at even larger values of $b$. Indeed, its phase takes a longer time to jump to different values, which demands impractically long simulations in order   
to see $\Tr(U)$ averaging out to zero. The phenomenon is not a phase transition because for finite $N$ there cannot be phase transitions and indeed at fixed $b$ the phenomenon is ameliorated when $N$ grows. It can be regarded as a phenomenon similar to a topology freezing. A genuine large $N$ phase transition of our model, replicating the one of the principal chiral model takes place at smaller values of $b$, and will be the subject of the next section. 

A final comment is that adding a second exponential one can fit all the correlation functions in the full range (including zero). The lowest mass is compatible with the values obtained before, and the additional mass is around 4 times bigger than this. The coupling to the state is around 0.1 times the corresponding one of the lower mass state.

\section{Analysis of the large $N$ phase transition}
\label{sect:6}
More than twenty years ago, Campostrini, Rossi and Vicari~\cite{Campostrini:1994ih} studied the large $N$ behavior of the $SU(N)$ lattice chiral model by Monte Carlo simulations with $N$ in the range $9 \le N \le 30$.  They found that the specific heat has a peak around $b=0.306$, whose  height  grows as $N$ increases, suggesting that there is a large $N$ second order phase transition. The reduced model allows us to reach much bigger values of $N$ so we can analyze the existence and nature of the phase transition in greater depth. The purpose of this section is to present the results of our study of the  TRPCM around $b=0.306$ with $N$ in the range $67 \le N \le 233$.

\begin{figure}
\centering
\includegraphics[width=90mm]{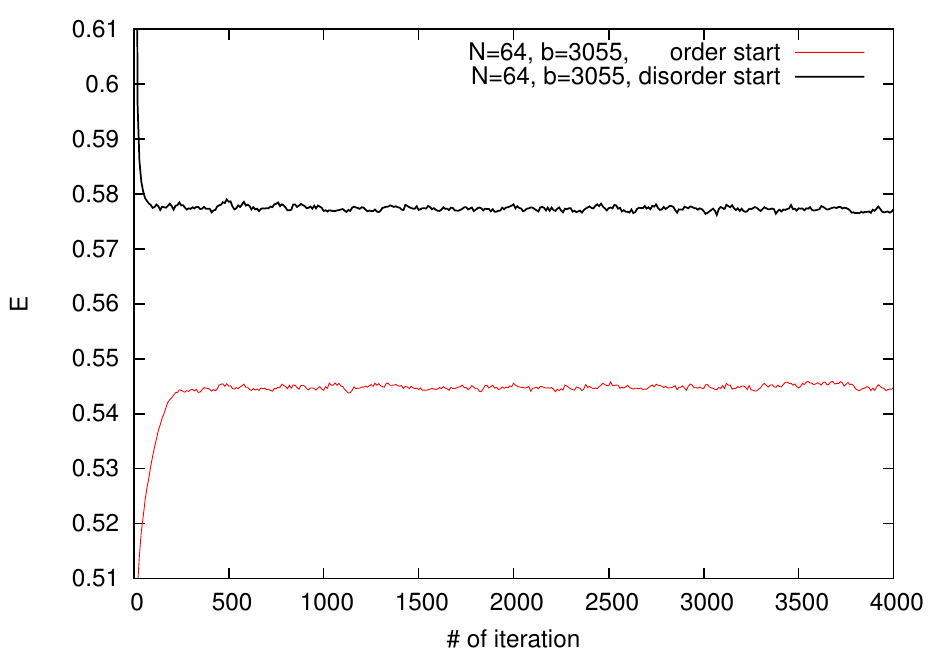}
\caption{Time history of the internal energy $E$ of the ordinary lattice chiral model.}
\label{fig:cm}
\end{figure}

Apart from studying the TRPCM we also explored the ordinary  lattice model on a $64\times64$ lattice at $b=0.3055$ and $N=64$. In fig. \ref{fig:cm}  we show the time history of the internal energy $E$ starting from both ordered and disordered configurations. The pattern shows two different values for each starting configuration, suggesting  that the phase transition is first order rather than second order. The amount of computer time required, due to the large number of degrees of freedom, limited the number of iterations of the model to 4000 in each case.

\begin{figure}
\centering
\includegraphics[width=90mm]{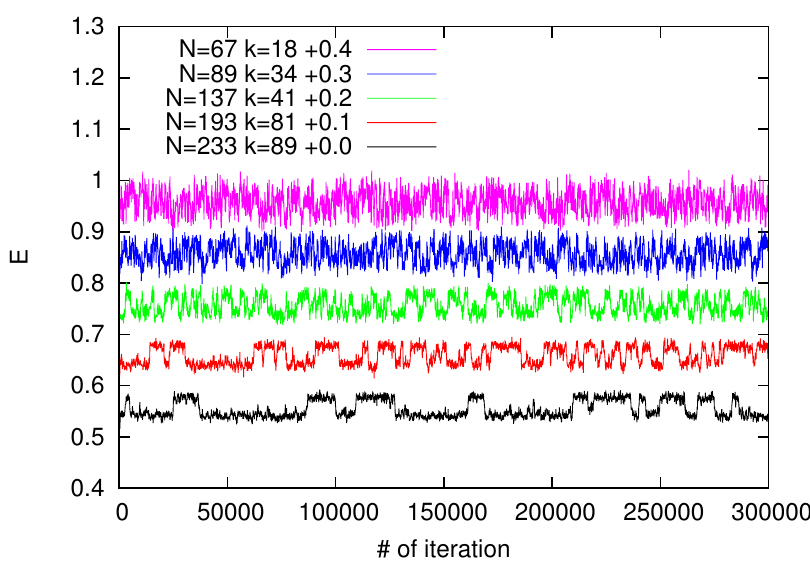}
\caption{Time history of the internal energy $E$ of the TRPCM.  Except $N=233$, the values of $E$ are  shifted upward to make the figure readable.}
\label{fig:trajectory}
\end{figure}

\begin{figure}
\centering
\includegraphics[width=90mm]{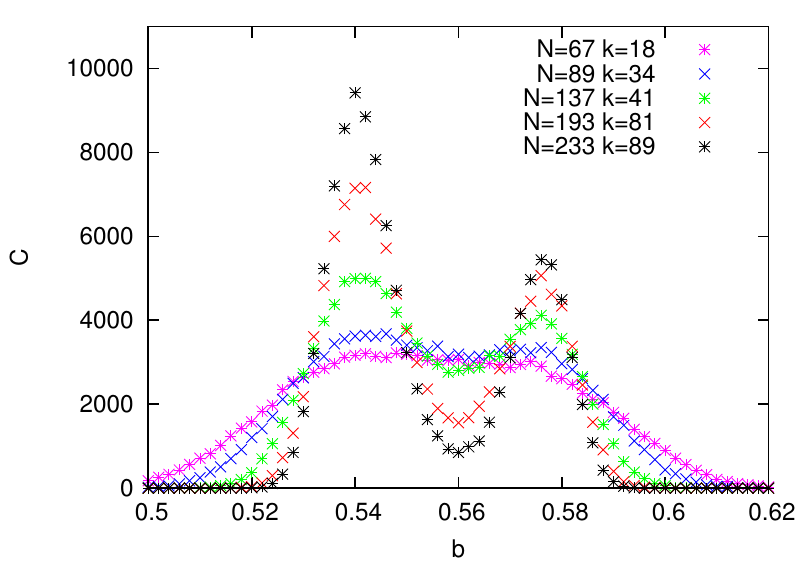}
\caption{Histogram of the internal energy $E$ of the TRPCM.}
\label{fig:histgram}
\end{figure}

At $b=0.306$ we have run $1.2\times 10^7$ Monte Carlo iterations for the  TRPCM with $N=67, 89, 137, 193$ and 233.  In Fig. \ref{fig:trajectory}, we show the time history of $E$ for the first 300000 iterations, plotting only the values every 100 iterations. The actual values have been displaced vertically to allow all values of $N$ to be put in the same plot. It is clear that at $N=67$, the  fluctuations of the internal energy are much bigger than those of the ordinary lattice chiral model with $N=64$ shown in Fig. \ref{fig:cm}.  This is expected since the Energy of the ordinary model is an average over the 
$64\times 64$ points of space. 

As we increase $N$ the dispersion of the internal energy of the TRPCM decreases and the existence of two different states  becomes clear.  This can be also seen by looking at the histogram of $E$ values displayed in fig. \ref{fig:histgram}. The  two peak structure 
becomes better defined  as $N$ increases, while the peak positions do not seem to change much. This is all pointing towards the first order character of the transition.

\begin{figure}
\centering
\includegraphics[width=90mm]{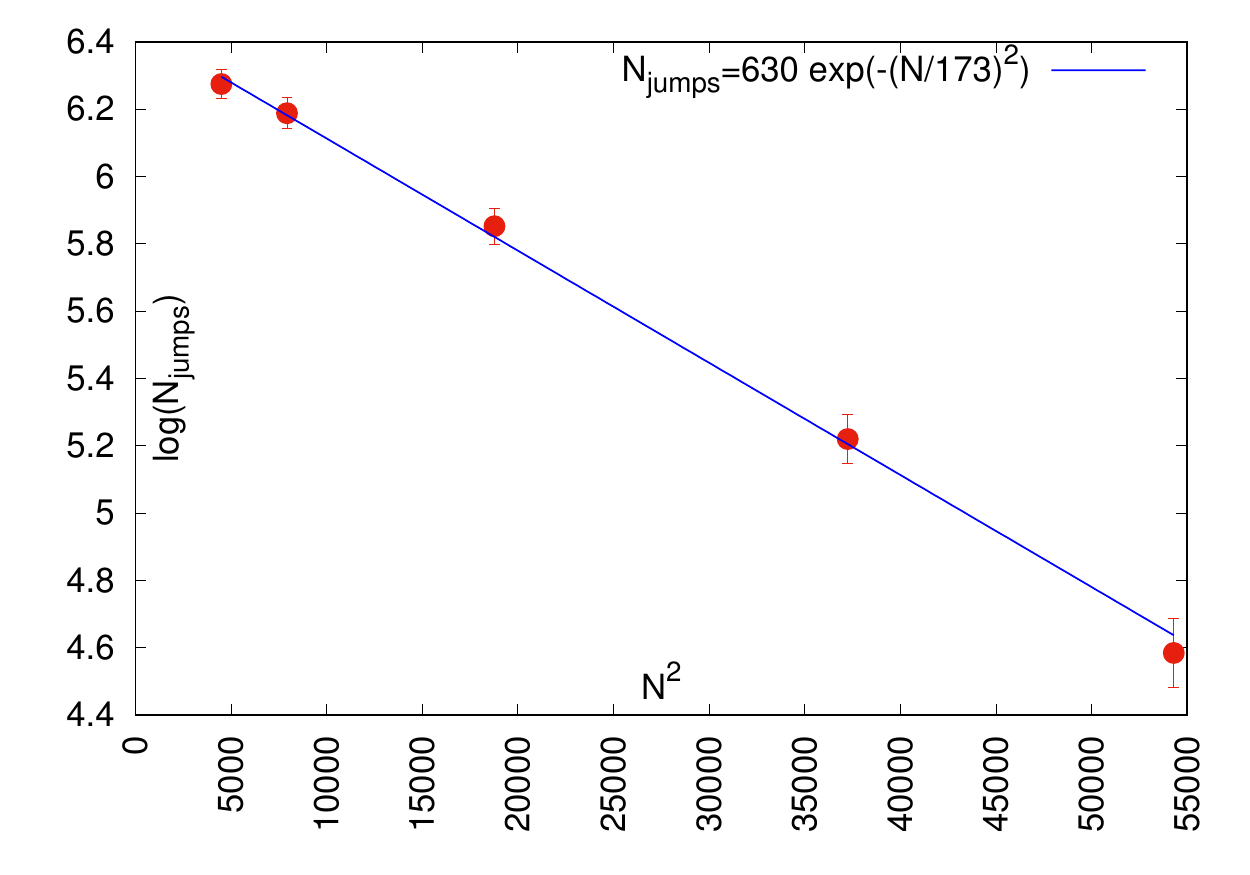}
\caption{The logarithm of the number of flip-flop jumps $N_{jumps}$ between the two phases as a function of $N^2$ .}
\label{fig:interval}
\end{figure}

It is also clear from Fig.~\ref{fig:trajectory} that the number of flip-flops between both states is decreasing as $N$ increases. To quantify this phenomenon, we smoothen out the fluctuations by averaging the measurements in groups of 10. Then we count the number of jumps from one phase to the other $N_{jumps}$. The logarithm of the values are displayed versus $N^2$ in fig. \ref{fig:interval}, together with a linear fit to the data. This implies that  the number of jumps decreases exponentially with the number of degrees of freedom of the system ($N^2$), or conversely that the tunneling times grow proportionally to the inverse of this number.  This is precisely the expected behaviour for a large $N$ phase transition. Incidentally, the values are consistent with the presence of no jumps in the 4000 iterations of the ordinary model at $N=64$. To make a final check we performed  $3\times 10^6$ iterations at $N=377$ with $k=233$ and found no flip-flops. 

\begin{figure}
\centering
\includegraphics[width=90mm]{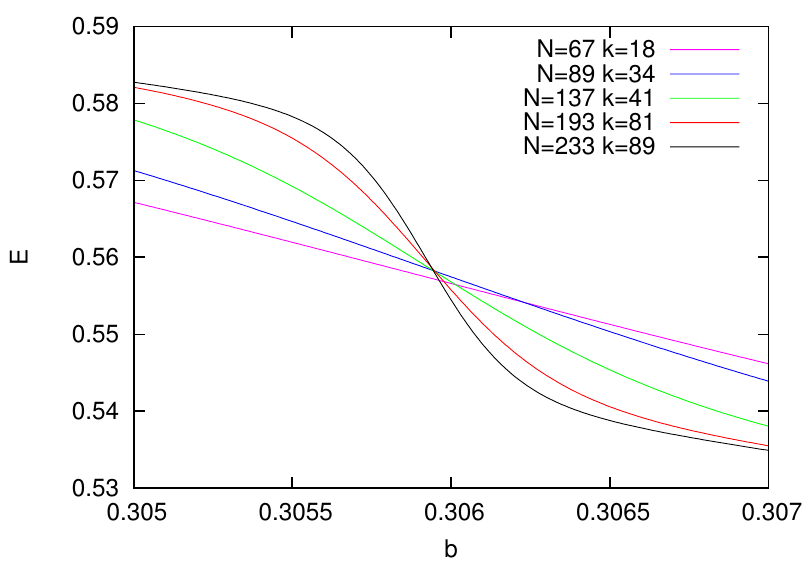}
\caption{$b$ dependence of the internal energy $E$ for $N=67, 89, 137, 193$ and 233.}
\label{fig:E}
\end{figure}

\begin{figure}
\centering
\includegraphics[width=90mm]{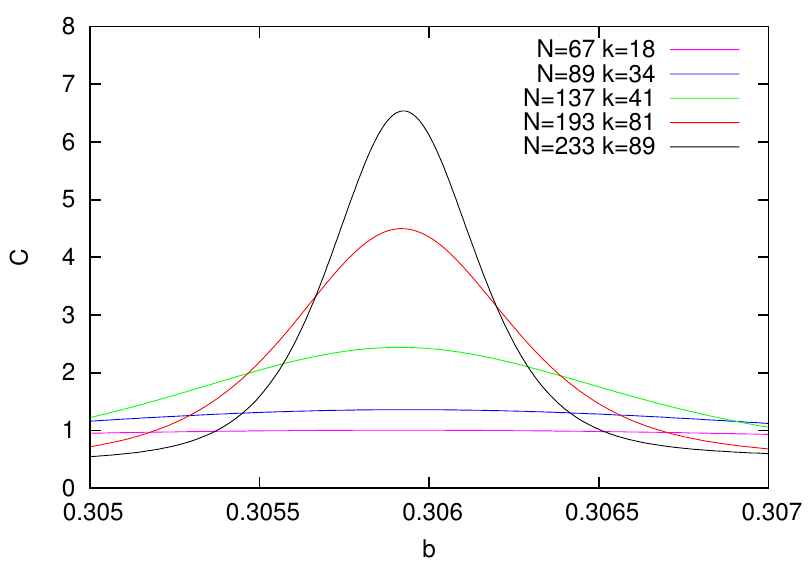}
\caption{$b$ dependence of the specific heat $C$ for $N=67, 89, 137, 193$ and 233.}
\label{fig:C}
\end{figure}

We also studied the $b$ dependence of $E$ and the specific heat $C=4b^2N^2(<E^2>-<E>^2)$ close to the phase transition by the re-weighting method.   Fig. \ref{fig:E} and \ref{fig:C} show the results of these analysis.  If the transition is second order, the peak height of C should behave as $C_{max}(N) \sim N^\alpha$ with $\alpha$ the critical index characterizing a second order phase transition.
However, $C_{max}(N)$ seems to be  approaching a growth with the exponent $\alpha=2$ characteristic of a first order phase transition. Actually using the peak values for the two largest values of $N$ (233 and 193) the value of $\alpha$ is found to be $1.985$. At the same time the profile for $E$ approaches a step function.

If the large $N$ phase transition were of second order, the correlation length should diverge  as we approach the phase transition. This is certainly not consistent with what we observe when coming from the ordered phase. From our data  at $N=233$ our measured  correlation lengths  for $b=0.307, 0.308, 0.309$ and $0.310$ are  (in lattice units) 
$3.82(2), 4.12(4), 4.45(10)$ and  $4.61(20)$ respectively. Errors include systematics. We also determined the correlation length at $b=0.306$ from the ordered data at $N=377$ being $3.64(8)$.

\begin{figure}
\centering
\begin{subfigure}[b]{0.5\textwidth}
\includegraphics[width=\textwidth]{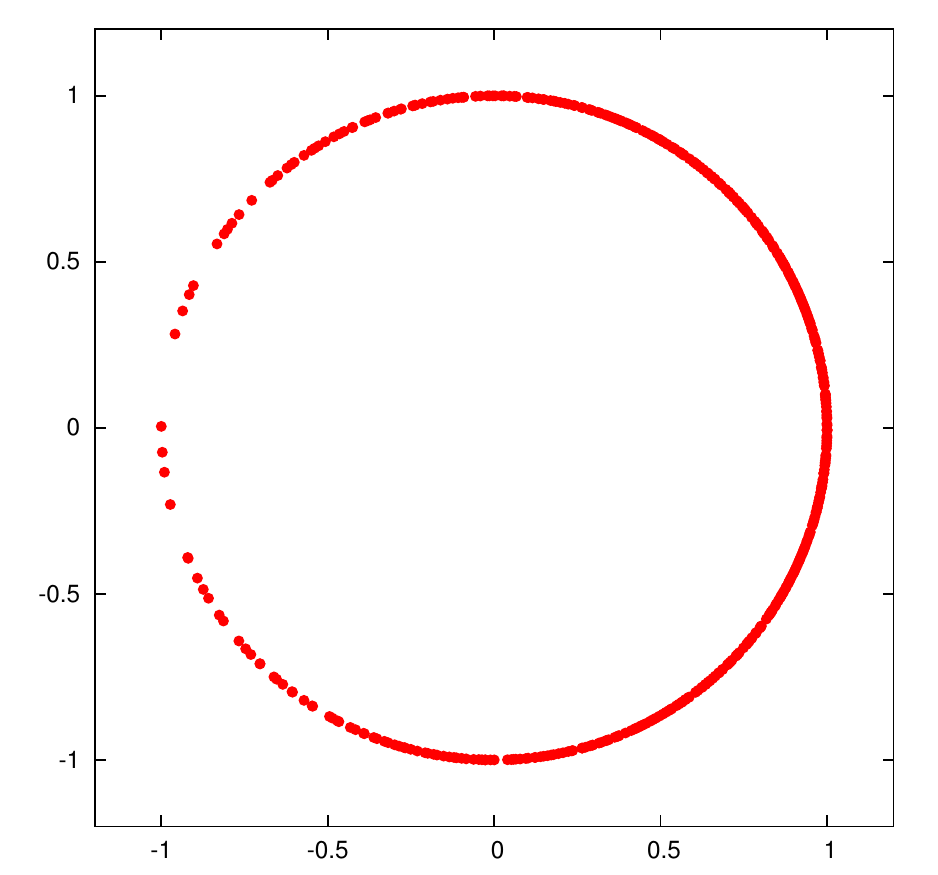}
\caption{ Disordered phase. }
\label{fig:eigena}
\end{subfigure}%
\begin{subfigure}[b]{0.5\textwidth}
\includegraphics[width=\textwidth] {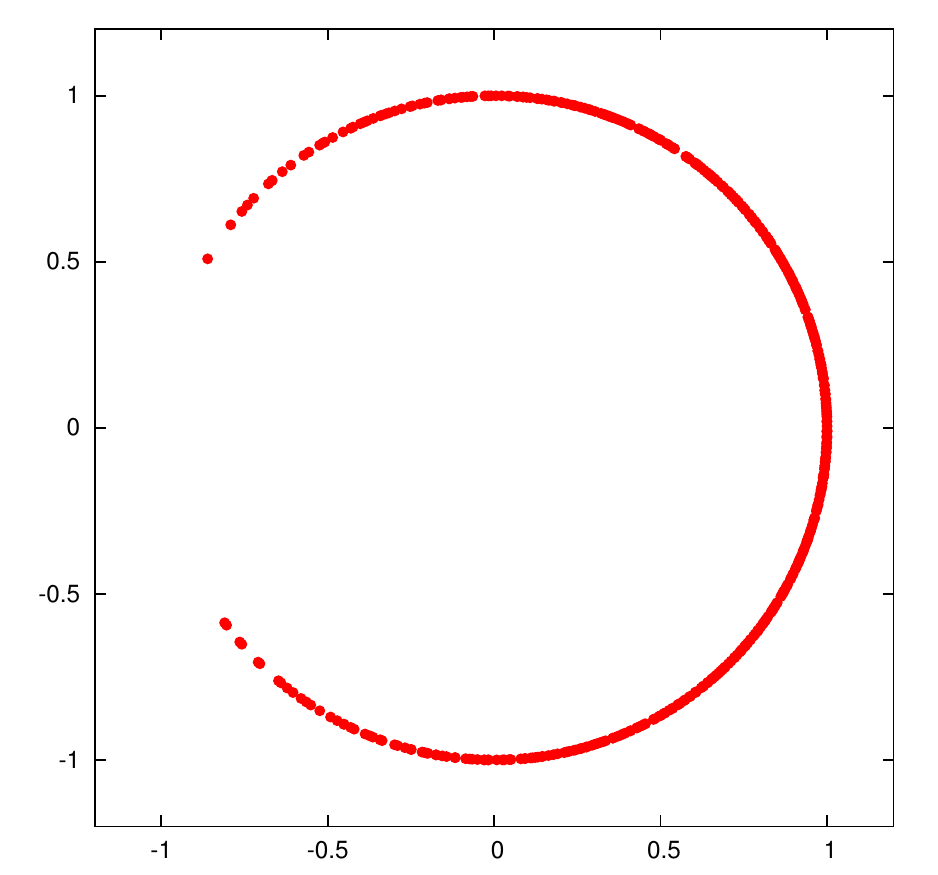}
\caption{Ordered phase.  }
\label{fig:eigenb}
\end{subfigure}%
\caption{Eigenvalue distribution of $U\Gamma_\mu U^\dagger \Gamma_\mu^\dagger$.}
\label{fig:eigen}
\end{figure}

Given that the transition is first order one might wonder about the different behaviour of observables in the two phases, apart from the value of the energy $E$. For that purpose we calculated the eigenvalue distribution of $U\Gamma_\mu U^\dagger \Gamma_\mu^\dagger$ in both ordered and disordered states.  To this end, we selected 30000th and 60000th $N=233$ configurations shown in 
Fig. \ref{fig:trajectory}.   The eigenvalue distribution are shown in Fig. \ref{fig:eigen}(a) for the disordered  30000th configuration and in Fig. \ref{fig:eigen}(b) for the ordered 60000th configuration.
It is clear that the eigenvalue distributions are quite different in the two phases.   
  
\section{Conclusions and Outlook}
\label{s.conclusions}
In this paper we have re-examined the  matrix model obtained by
applying the twisted reduction procedure to the principal chiral
model. Earlier, the two-dimensional model was found to give controversial
results in its claim to be equivalent to the ordinary model at large
$N$. Due to the  new information acquired lately about similar
issues of the reduced gauge theory model, we have been able to show
that the difficulties can be avoided by choosing adequately the
integer flux appearing in twisted boundary conditions. The criteria
are very much the same as the ones found in gauge theories. Applying
these guidelines we were able to show with great numerical precision
that the results of the reduced and ordinary models agree within
errors. Concerning the finite $N$ corrections its size also gets
reduced by the choice of flux. An even stronger reduction is obtained
by using only a partial reduction~\cite{Narayanan:2003fc}-\cite{GonzalezArroyo:2005dz} to a twisted $2\times 2$ 
or $3\times 3$ box. It is beyond any doubt that using twisted boundary conditions in the study of this model pays off. 

Our test of the reduced model has been extended to the continuum limit
where it can be directly tested against exact results obtained for the principal chiral model. Our results also give conclusive evidence of scaling as
dictated by the beta function of the model. 

Finally, we have used the reduced model to investigate the interesting
issue of the large $N$ phase transition of the principal chiral model. The larger values of $N$ that can be reached are an important benefit when studying this problem. Our conclusion is that the transition is actually of first order nature. 

The simplicity of the TRPCM (a single unitary matrix model) makes it suitable for future studies including attempts of an exact solution. Several open problems, already mentioned in the introduction, could be analyzed.

%\acknowledgments
\section*{Acknowledgments}
During the long time span of this research A.G-A has benefitted from
many interesting discussions with many colleagues. Here we want to
signal out specially Margarita García Pérez, Herbert Neuberger, Gerald
Dunne, Rajamani Narayanan, Peter Orland and Oleg Evnin. Special thanks go also 
to the members of the Department of Theoretical Physics of 
Tata Institute of Fundamental Research (Mumbai) for their warm hospitality and 
the many exciting scientific exchanges.
Finally, A.G-A also wants to thank the hospitality of the Department of Physics 
at Rutgers University where the last steps of this work were completed.  
A.G-A acknowledges financial support from the MINECO/FEDER grant
FPA2015-68541-P  and the MINECO Centro de Excelencia Severo Ochoa
Programs SEV-2012-0249 and SEV-2016-0597, as well as CUniverse
research promotion project by Chulalongkorn University (grant
reference CUAASC) for funding his stay in Thailand and the Ministry of
Education of Spain (programa Salvador de Madariaga) for funding of his stay at Rutgers.
M. O. is supported by the Japanese MEXT grant No 17K05417 and the MEXT program for promoting the enhancement of research universities. 
We acknowledge the use of HITAC SR16000 computer at KEK and the Hydra cluster at IFT.

\bibliography{apclas}
\end{document}